\documentclass[aps,prl,preprint,groupedaddress,showpacs]{revtex4}

\usepackage{graphicx}
\begin{document}


\title{Network as a Complex System: Information Flow Analysis }


\author{Vladimir Gudkov }
\email[]{gudkov@sc.edu}
\affiliation{Department of Physics and Astronomy
\\ University of South Carolina \\
Columbia, SC 29208 }
\author{Joseph E. Johnson}
\email[]{jjohnson@sc.edu}
 \affiliation{Department of Physics and
Astronomy
\\ University of South Carolina \\
Columbia, SC 29208 }


\date{\today}

\begin{abstract}
A new approach for the analysis of information flow on a network
is suggested using protocol parameters encapsulated in the package
headers as functions of time. The minimal number of independent
parameters for a complete description of the information flow
(phase space dimension of the information flow) is found to be
about 10 - 12.
\end{abstract}

\pacs{89.20.-a, 89.70.+c, 89.75.-k, 05.45.Gg}

\maketitle


Both local and wide area (internet) networks are now intensively
used for different kinds of communication and information
exchange. A variety of different hardware devices from low level
switches and routers to PC's and to super-computers are involved
in this process. The diversity of software to support this process
even more impressive including many operating systems (with
possible variations of open code ones) and application software.
To control the information exchange, many protocols (like
currently popular TCP/IT one) have been created. Different
protocols (which contain a number of important parameters) are in
charge of different processes during information exchange. Even to
establish a host-to-host connection between two computers, one
needs a multiple exchange of information.

Therefore, information exchange is a very complicated process
requiring a careful analysis in order to support
 network simulations and the optimization of the
exchange. The understanding of information flow on the network is
necessary for real time network monitoring and for the difficult
problem of  detection\cite{indet} and prevention of network
intrusions in real time. Possible solutions may be much more
efficient provided we understand network traffic in detail.

In this letter we consider some problems related to the general
properties of information flow on a network. Since, the process of
information exchange is extremely complicated, we apply some
approaches for analysis of complex systems in physics to the
information flow on networks. The first question we would like to
address is: how many parameters do we need to describe the
information flow on the network? In other words, what is the
dimension of the network parameter space? It is also important to
know the extent and manner of how this dimension depends on the
network topology, its size, and the operating systems involved in
the network.

To define the fundamental objects related to information flow we
recall that network information is transferred by packages of
restricted size which are exchanged between computers. The
structure of the packages vary from one to another and depend on
the transfer process. In general each package consists of a header
and encapsulated data. The header consists of encapsulated
protocols related to different layers of communications: from a
link layer to an application layer. In this letter we will ignore
 the encapsulated data, since it does not affect the
package propagation through the network. Rather, we are interested
in the information contained in the header which controls all
network traffic.

Using these packages as fundamental objects for information
exchange we apply some methods known in physics as tools for the
analysis of complex ergodic systems to the network traffic
analysis.

The possibility of reconstructing the dynamics of ergodic systems
from an experimental signal (see, for example ref.\cite{chaos}and
references therein) is related to the Ma\~{n}\'{e} theorem
\cite{mane} and can be briefly stated as follows: If an attractor
of the system has finite Hausdorff dimension, then using a
sufficient number of variables (less than about twice of the
Hausdorff dimension) it is possible to plot trajectories of the
attractor in the space of the variables without a self-crossings
of these trajectories.

To define the dimension of the information flow on a network we
use the approach for analysis of observed chaotic data in physical
systems as suggested in papers\cite{abar1,abar2}(see, also
references therein). Any dynamical system with dimension $N$ can
be described by the system of $N$ differential equations of the
second order in configuration space or by the system of $2N$
differential equations of first order in phase space. We assume
that the information flow can be described in terms of ordinary
differential equations (or discrete-time evolution rules) for some
unknown functions $\overrightarrow{F}(\overrightarrow{g})$ in a
(parametric) phase space $\overrightarrow{g}$
\begin{equation}
\frac{d\overrightarrow{g}(t)}{dt}=\overrightarrow{F}(\overrightarrow{g}(t)).
\label{sde}
\end{equation}
Here we use for the sake of simplicity, a continuous
representation for dynamical systems. The discrete representation
does not change any results.  (For discussions of well-known
relations between continuous and discrete representations see,
e.g. refs.\cite{chaos,abar1}.) We do not know the right dynamical
variables $\overrightarrow{g}$ which describe the motion
(development in time) of our system in $N$-dimensional phase
space. However, we assume that they could be related to some
chosen $\overrightarrow{x}$ parameter representation as
$\overrightarrow{g}(\overrightarrow{x}(t))$. (Note that the
dimensions of vectors $\overrightarrow{g}$ and
$\overrightarrow{x}$ are different, in general.) Let us assume
that we measure a scalar quantity $s$ which is a function of these
dynamical variables
$s(\overrightarrow{g}(\overrightarrow{x}(t)))=s(t)$. To extract
the dimension of the system phase space from the time-dependance
of the variable $s$ we construct $d$-dimensional vectors for all
possible  $n$
\begin{equation}
y^d(n)=[s(n),s(n+T),s(n+2T),\ldots , s(n+T(d-1))] \label{yvec}
\end{equation}
from values of the parameter $s$ at equal-distant time intervals
$T$: $s(t) \rightarrow s(T\cdot n) \equiv s(n)$, where $n$ is an
integer numerating $s$ values at different times. Building sets of
vectors $y^d(n)$ for spaces with increasing dimension $d$, we
calculate the number of nearest neighbors $\nu ^d(n) $ for each
point represented by the vector $y^d(n)$. Imagining that the
vector $\overrightarrow{y}^d$ corresponds to the system trajectory
in $d$-dimensional space we are looking for that dimension for
which the trajectory will not intersect itself. For the discrete
type of
 trajectory represented in terms of vectors
$\overrightarrow{y}$, the intersection means the existence of a
number of nearest neighbors in the vicinity of the intersection
decreases with increasing  dimension $d$ of the parametric space.
Therefore, increasing the dimension $d$ step-by-step and plotting
the number of false nearest neighbors(FNN), one can reach a point
with no FNN. This is because these (false) neighbors arise from
the projection of far away parts of the trajectory in higher
dimensional space and at the point where there are no FNN we come
to the upper limit of the parametric space for our system.
 This could be illustrated by the
simple example for a two-dimensional circle. If we project the
circle on a one-dimensional space, we get an interval with two
degenerate points along the projection axis. Increasing the
dimension by 1 we come to the original two-dimensional circle
without the degeneracy. Thus, the degenerate points in 1-dimension
which have moved to the opposite sides of the circle in
2-dimensions could be called a false nearest neighbors (FNN). By
unfolding the space further, to 3-dimension3, or further, one will
no longer get a false nearest neighbors since a higher dimensional
space covers the two-dimensional space to which the circle belongs
to.

Therefore, if the parameter $s$ is sensitive to the system
dynamics and if it has been measured for a long enough period of
time, provided the time interval $T$ was chosen properly, than the
number of false nearest neighbors decreases with the increasing of
the dimension $d$ up to some limit which corresponds to the real
dimension of the system under consideration\cite{abar1,chaos}.

To give a general idea why this algorithm is related to the
dimensionality of the dynamical system governed by Eq.(\ref{sde}),
 we recall a simple method for numerical solution
of a system of differential equations of $N$-th order based on
finite differences approach. First of all one can rewrite the
system of $2N$ differential equations of the first order
(\ref{sde}) as a differential equation of $2N$-th order for the
scalar function $s(t)$. To solve it one needs to calculate the
function $s(t)$ and all its derivatives up to order $2N$. One can
see from the simplest numerical expressions for derivatives of the
function $s(t)$

\begin{eqnarray}
\frac{ds(t)}{dt}\simeq \frac{s(t+T)-s(t)}{T}  \nonumber \\
\frac{d^2s(t)}{d^2t}\simeq \frac{s(t+2T)-2s(t+T)+s(t)}{2T^2}
\label{deriv} \\
\ldots \hskip1cm \ldots \hskip1cm \ldots \hskip1cm \ldots
\hskip1cm \ldots \nonumber
\end{eqnarray}
that the knowledge of the derivatives leads to the knowledge of
the set of parameters $s(i)$ on the proper chosen period of time
for a sufficiently small interval $T$. This set of parameters
$s(i)$ corresponds exactly to the vector $\overrightarrow{y}^d(n)$
in Eq.(\ref{yvec}).

To analyze information flow on a network we use ``tcpdump''
utilities, developed with the standard of LBNL's Network Research
Group \cite{tcpd}, to monitor local network communications. Also
we use tcpdump files for DARPA Intrusion Detection Evaluation from
MIT Lincoln Laboratory. It is important for our research that we
did not discard {\it apriori} any information contained in the
binary tcpdump file in order to be able analyze any parameters in
the header of packages transferred trough the network.
Technically, we use  decoding software developed by our research
team to extract any combination of the header parameters of the
packages (including Ethernet and IP protocols as well as transport
layer protocols like, TCP, UDP and ICMP) as functions of time. It
should be noted that the procedure of network monitoring allows us
to collect information (to dump files) at different points in the
networks and to separate incoming, outgoing and internal network
traffics.

First at all, we observed that the protocol parameters for
host-to-host communication (i.e all information using in
protocols) could be divided into two separate groups with respect
to the preservation or change in value during the package
propagation through the network. We call these two groups of
parameters dynamic and static. The dynamic parameters  may change
during the package propagation through the network (internet). For
example, a ``physical'' address of a computer, which is  the MAC
parameter of the Ethernet protocol, is a dynamic parameter because
it can change if the package has been re-directed by a router.
Conversely, the source IP address is an example of static
parameter because it does not change its value during a packet
propagation. For the purposes of information flow analysis, it is
reasonable to use only static parameters
 since they may carry intrinsic properties of the
information flow neglecting the network (internet) structure. From
the other hand if we were interested in the influence of the
network structure on the packages propagation, the dynamic
parameters could be a choice. However, this is out of the scope of
this letter.

\begin{figure}
\includegraphics{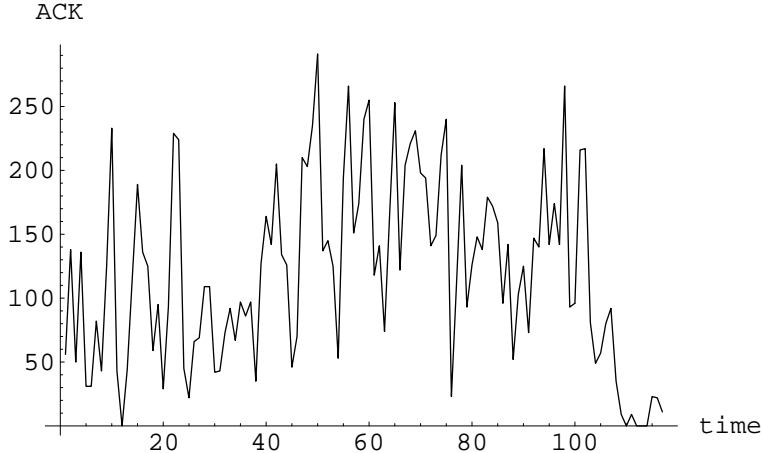}
\caption{ACK flag in TCP/IP protocol as a function of time (in
$\tau = 5 sec$ units).} \label{fig:flag}
\end{figure}

\begin{figure}
\includegraphics{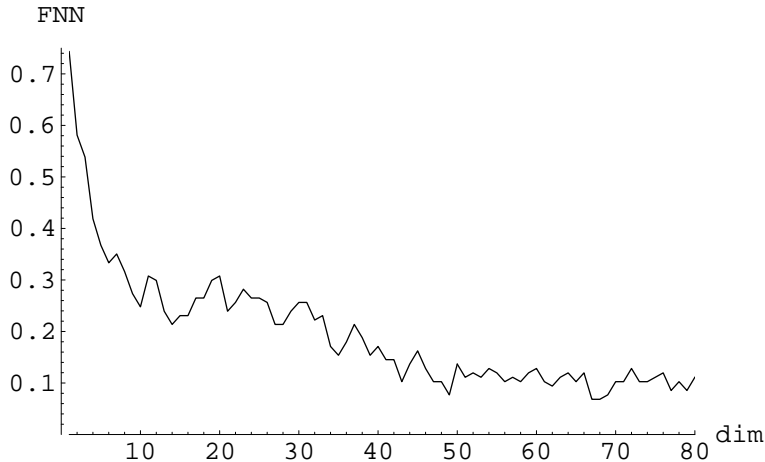}
\caption{Relative number of false nearest neighbors as a function
of dimension of unfolded space.} \label{fig:fnn}
\end{figure}

\begin{figure}
\includegraphics{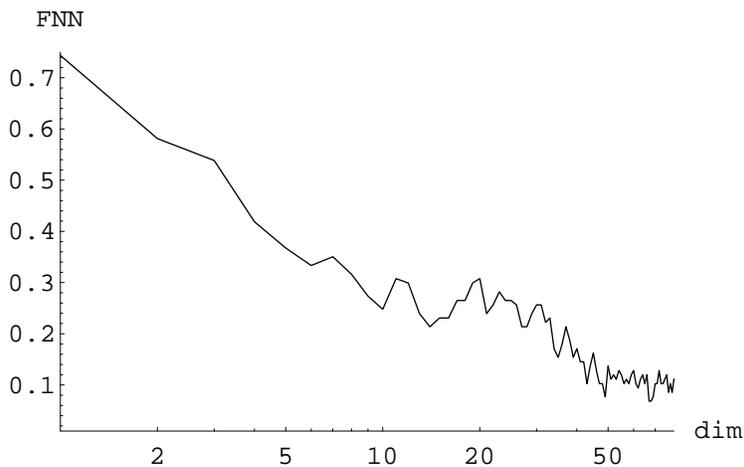}
\caption{Logarithm of relative number of false nearest neighbors
as a function of dimension of unfolded space.} \label{fig:lfnn}
\end{figure}

Now we can apply the above approach to understand the scale of
dimension of the information flow and its dependance on network
structure. We have done the analysis with randomly chosen static
parameters from the protocols in headers of the packages
travelling on networks. To illustrate the procedure, let us
consider an example for one particular parameter, the density
(number of appearance in the fixed interval of time  $\tau$) of
the ACK flag for TCP/IP protocol. The density of the ACK flag as a
function of time has been extracted from the measurement (tcpdump
file) on the gate of the local network with about 50 computers
running under Windows 2000, Windows NT, Linux and Unix operating
systems. This function is shown in Fig.\ref{fig:flag}.
It is important that this parameter (like any other parameters
from the protocol) can be represented in the form of
time-dependant function convenient for numerical analysis.
Therefore, we can apply the algorithm for the restoration of the
dimension of dynamical systems for determination of the dimension
of information flow using this ACK parameter as a scalar function
$s(t)$ related to the information flow dynamics. As a result, it
gives us the dependance of the relative number of false nearest
neighbors (FNN) as a function of dimensionality of the unfolding
parametric space (see Fig.\ref{fig:fnn} and Fig.\ref{fig:lfnn}).
One can see that the number of false nearest neighbors rapidly
 decreases  up to about dimension's value 10 or 12. After that
 it shows a slow dependance, if any at all, on the
dimension. To define the embedded dimension of the information
flow on the base of FNN analysis one can use different existing
methods. There is the possibility to discriminate real FNN
calculated  distances of false nearest neighbors and define a
threshold - the maximal allowed value for the minimal distances -
to count a nearest point as an neighbor (see, e.g.
ref.\cite{abar1}). Another method \cite{muentr} involves the
preliminary optimal choice of the time lag based on the mutual
entropy calculation for the signal (in our case the parameter
ACK). For the proper analysis, a combination of these methods
should be applied. Since we are interested not in the precise
Hausdorff dimension of the system (which is probably a fractional
one) but rather in the estimation of the number of linear
independent parameters for the complete description information
flow, we will provide here another criteria which helps us to
understand the general features in the behavior of FNN as a
function of dimension in Figs. (\ref{fig:fnn}) and
(\ref{fig:lfnn}). We use as a guideline the
observation\cite{rmt1,rmt2} that, while a system changes its
behavior from integrable to ergodic one, the distribution of the
spacing between nearest neighbors of the system eigenvalues moves
from a Poisson to Wigner-type distribution.  It should be noted
that we are taking into account the analogy between a
non-correlated Poisson distribution and a Wigner-type distribution
with long range correlations for the eigenvalues of the system and
for the false nearest neighbors in the phase space but we do not
assume a relation between eigenvalues and false nearest neighbors
themselves.

\begin{figure}
\includegraphics{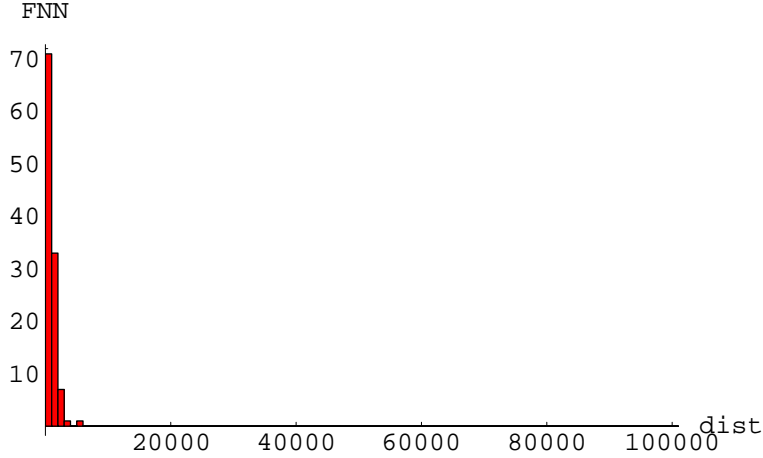}
\caption{Distribution of points under suspicion to be false
neighbors for 2-dimensional space.} \label{fig:h2}
\end{figure}

\begin{figure}
\includegraphics{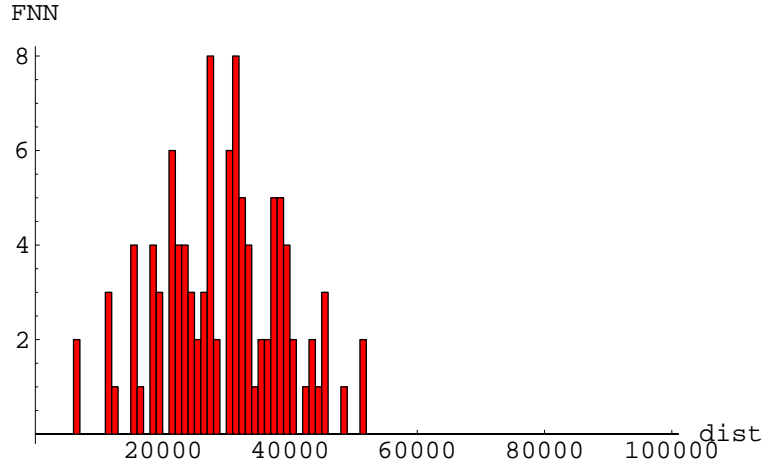}
\caption{Distribution of points under suspicion to be false
neighbors for 11-dimensional space.} \label{fig:h11}
\end{figure}

From this point of view the distribution of the number of false
nearest neighbors
 as a function of their distances from the points
under consideration will depend on the nature of the system
behavior. If the false nearest neighbors originate from the
projection of a high dimensional system attractor onto a lower
dimensional manifold, than one expects randomly independent
Poisson distribution of their numbers as a function of distances
in the vicinity of the points of under consideration.  On the
other hand, if the false nearest neighbors originate from chaotic
motion due to white noise in the system itself or in surrounding
environment (as it may be in our case), one can expect a
Wigner-type distribution. To be correct, in the last case the term
`` false nearest neighbors'' is not a relevant one anymore since
the distributions of these neighbors are governed not by the
system itself but rather by a noise from the environment. Using
this criterion one helps understand the nature of the tails for
FNN dependencies on Figs.(\ref{fig:fnn}) and (\ref{fig:lfnn}).
For our case the FNN distributions are shown for dimensions 2 and
11 on Figs. (\ref{fig:h2}) and (\ref{fig:h11}).  On the Fig.
(\ref{fig:h2}) one can see a Poison-like distribution for small
($d=2$) dimension of the unfolded space. This distribution
continuously transforms into Wigner-like one when the dimension of
the unfolded space approaches 11 (see Fig.(\ref{fig:h11})). It
should be noted that our criteria is consistent with the algorithm
considered in ref.\cite{abar1} since the Wigner-like distribution
appears at large distances which are excluded from the region of
definition of false neighbors.

Therefore, one can conclude that the above analysis, within the
given accuracy, shows that the information flow on the network can
be described in a parametric (phase) space with dimension of about
10 to 12, or in other words the information flow dynamics can be
described in terms of 12 (or less) independent parameters.

Strictly speaking to make such conclusion we need to know that the
chosen parameter (ACK, in the given example) has strong relation
to the dynamics of the information flow and  that our choice of
time lags ($\tau = 5sec$ in the above case) is adequate for the
time scale of the system. We cannot prove the relation of the
parameter (ACK) to the system dynamics since there is no a
reasonable theory to describe the information flow on the network.
What we have done instead is the similar analysis of other (about
10) parameters from IP and TCP protocols. Within the same accuracy
($\pm 1$ degree of dimension) we have obtained the result that the
dimension of the information flow extracted using different
parameters does not depend on the parameter choice and has the
same value of 10 to 12. To address the question about the choice
of the time lags we changed the parameter $T$ from the scale of
fractions of seconds to hundreds of seconds (where it was
possible). An expected dependance has been observed: by increasing
(decreasing) the $T$, one could decrease (increase) the chaotic
(white noise, or high dimensional) part of distribution shown on
Figs. (\ref{fig:fnn}) and (\ref{fig:lfnn}) for large (small)
values of the leg $T$ with relatively independence of $T$ in the
wide region around the optimal value.

The next step of our analysis was the study of possible dependency
of the obtained dimension of the information flow from network
structure: its size, topology, and working operating systems. We
used tcpdump files for networks with sizes from several computers
up to more than hundred, separated internal and external flows,
and used networks with different operating systems: Windows NT,
Windows 2000, Linux and Unix. For all these cases the obtained
dimension of the information flow was consistent with the value of
10 - 12.

This gives us reason to conclude that the information flow on the
network is decoupled from network structure and can be considered
as an independent (or almost independent) value with its dynamics
described as a trajectory in a phase space with the dimension
about 10 - 12. We can consider a rough analogy for the information
flow on network with a liquid flow through the systems of pipes.
In the last case one can use the Navier-Stokes equation. For the
information flow the corresponding equation is unknown. However,
it is important to know that one needs not more than 12 parameters
to describe it.

\begin{acknowledgments}
We thank staff of Advanced Solutions Group for technical support.
 This work was supported by the DARPA Information Assurance and
 Survivability Program and is administered by the USAF Air Force
 Research Laboratory via grant F30602-99-2-0513, as modified.
\end{acknowledgments}

\end{document}